\documentclass[amssymb,amsmath,nobibnotes,aps,prl,preprint,groupedaddress,showpacs]{revtex4}

\usepackage{graphicx}

\begin{document}


\title{Interference of an array of atom lasers}

\author{Giovanni Cennini}
\author{Carsten Geckeler}
\author{Gunnar Ritt}
\author{Martin Weitz}
\affiliation{Physikalisches Institut der Universit\"{a}t T\"{u}bingen, Auf der
Morgenstelle 14, 72076 T\"{u}bingen, Germany}

\date{\today}

\begin{abstract}
We report on the observation of interference of a series of atom
lasers. A comb-like array of coherent atomic beams is generated by
outcoupling atoms from distinct Bose-Einstein condensates confined
in the independent sites of a mesoscopic optical lattice. The
observed interference signal arises from the spatial beating of
the overlapped atom laser beams, which is sampled over a vertical 
region corresponding to 2\,ms of free fall time. The average relative de
Broglie frequency of the atom lasers was measured.
\end{abstract}


\maketitle

Soon after the realization of optical lasers, the temporal
interference signal among two such sources was observed~\cite{1}.
This enabled investigations of the difference frequency and the
frequency stability of the optical sources. To date, advances in
the field of synthesizing and controlling optical frequencies
allow for the measurement and comparison of optical frequencies
with radiofrequency precision~\cite{2,3}. Atom lasers are coherent
atom sources whose outcoupled wave can be seen as a matter wave
analog to the radiation emitted by optical
lasers~\cite{4,5,6,7,8}.

We report on the measurement of the relative frequency of several
distinct atom laser beams. In the sites of a mesoscopic optical
lattice potential, rubidium atoms are initially cooled to quantum
degeneracy to form independent microcondensates. By smoothly
ramping down the confining lattice potential, the condensates are
coupled out into a comb-like array of independent, coherent atomic
beams directed downwards along the earth's gravitational
acceleration. Due to their divergence, the atom laser beams soon
overlap. The resulting interference pattern, as observed by
absorption imaging, gives a streak-camera like image of the
beating of the coherent matter wave beams. The temporal evolution
of the atom lasers relative phase was monitored over a vertical
range corresponding to 2\,ms of free fall time. The average
difference frequency between adjacent atom laser sources was in a
proof of principle experiment measured to an accuracy of 46\,Hz.

Bose-Einstein condensates are coherent ensembles of particles, all
populating a single one-particle state~\cite{9}. The coherence of
such quantum degenerate samples has first been verified in
interference experiments with two atomic Bose-Einstein
condensates, leading to the conclusion that each condensate upon
measurement can be described by a single, macroscopic quantum
phase~\cite{10}. Recently, the relative phase of two Bose-Einstein
condensates was continuously sampled by light scattering off the
internal atomic structure~\cite{11}. On the other hand, light
scattering can also imprint an atom phase~\cite{12}. Bose-Einstein
condensated atoms can be outcoupled from their confining
potentials to form atom laser beams~\cite{4,5,6,7,8}. By splitting
up and recombining a single atom laser beam, its spectral line
width was determined~\cite{13}.

Our experiment is based on the technique of direct generation of
Bose-Einstein condensates in optical dipole traps~\cite{14,15,16,8}.
Initially, cold thermal rubidium atoms ($^{87}\rm{Rb}$) are confined in
the antinodes of a one-dimensional optical standing wave generated
using mid-infrared radiation derived from a CO$_2$-laser.
Neighbouring lattice sites are spaced by $d =
\lambda_{\mathrm{CO_2}}/2 \approx 5.3\,\mu\mathrm{m}$. For our
experimental parameters, atom tunnelling between adjacent sites is
negligible, which guarantees that the atom lasers operate truly
independently. By applying a magnetic
field gradient of typically $10\,\mathrm{G/cm}$, we remove atoms in
magnetic field sensitive spin projections, yielding
an ensemble of atoms all populating the $m_F=0$
component of the lowest hyperfine ground state ($F=1$).
The thermal atoms are cooled evaporatively to quantum degeneracy
by lowering the trapping laser beam power
to produce an array of independent disk-shaped microcondensates,
as described more in detail in an earlier work
~\cite{17}. At the end of the evaporation stage, the CO$_2$-laser beam power
is 40 mW on a $30\,\mu\mathrm{m}$ beam radius.
We generate an array of in average seven
disk-shaped $m_F = 0$ microcondensates. The total number of atoms in this
array is 2000. The atom laser beams are generated by subsequently smoothly
ramping the CO$_2$-laser beam intensity towards zero in a 30\,ms long
ramp. The outcoupling occurs once the dipole trapping force does
not anymore sustain the atoms against gravity. In this way, a
comb-like structure of parallel atom laser beams directed
downwards is produced. Due to the use of $m_F = 0$ condensates in our all-optical
technique, fluctuations of the chemical potential due to stray
magnetic fields are suppressed to within the
$14\,\mathrm{fK/(mG)^2}$ quadratic Zeeman shift of this
clock-state~\cite{8}.

We note that the spatial phase evolution of each atom laser is
determined by the de Broglie wavelength $\lambda = h/p$ with
$p^2/2m = \hbar\omega-mgz$. One expects $\lambda$ to decrease
during the free fall as the momentum $p$ increases. Despite the
strong chirping of the spatial phase, monochromaticity of the
matter waves is clearly defined~\cite{18,19,20}. In contrast, a
time-dependent, fluctuating value of the chemical potential of the
condensates would lead to deviations from merely gravitationally
chirped matter waves, and relative variations can be measured in
interference experiments with other atom laser beams. More
generally, such a comparison can also reveal other sources of
differences in the independent beams frequency, such as variations
in the vertical trap position, which will lead to a gravitational
phase difference, or possibly even tiny differences in the atoms
mass.

A scheme of the potential experienced by atoms during the
onset of the atom laser beams operation is shown in
Fig.~\ref{fig:scheme}. Before the condensates outcoupling, the
trap vibrational frequencies are 1.9\,kHz and 150\,Hz along the
longitudinal and the radial trapping directions of the lattice
respectively. The atom laser beams from adjacent sites are
estimated to start overlapping at a time of $500\,\mu\mathrm{s}$
after outcoupling, and after 4 ms a complete overlap of all beams
occurs.

We experimentally observe the far-field interference pattern of
the atom lasers using absorption imaging. Fig.~\ref{fig:pattern}a
shows a typical interference pattern for an average of 7 populated
sites in the lattice, which equals the number of interfering atom
laser beams. The image gives a streak-camera like recording of the
interference pattern, in which the vertical position z relates to
the fall time via the ballistic free fall formula.
Fig.~\ref{fig:pattern}b shows a transverse profile of the
interference pattern averaged over a vertical length of
$44\,\mu\mathrm{m}$.

Let us next discuss the expected interference signal for
such an array of overlapping atom laser beams. We expect that the
macroscopic wavefunction of each atom laser beam is given by an
eigensolution of the Gross-Pitaevskii equation in the gravitational
field. As the total number of
atoms emitted from each source is fixed, we do not expect the atom
lasers to have a well-defined phase. However, upon measurement of
the interference pattern of the overlapping beams a projection
upon a state with well-defined relative phase is performed.

The optical lattice is in our experiment aligned orthogonally to
the axis of gravity. Let us denote the macroscopic condensate
wavefunction at
the $n$-th site ($n = 1,2,\ldots,N$) centered at position
$\mathbf{r}_n$ at time $t = 0$ as $a_n e^{i\theta_n}$, where $a_n$
gives the amplitude and $\theta_n$ the phase, the latter being
random in each realization of the experiment. After outcoupling,
the total wavefunction of the overlapping beams at position
$\mathbf{r}$ and time $t$ (with $t > 0$) can be formulated using
path integrals evaluated along classical space-time
trajectories~\cite{21}:
  \begin{eqnarray}
    \psi(\mathbf{r}, t) = \sum_{n=1}^{N} \textstyle
    a_n e^{i\theta_n} C(\mathbf{p}, t_{\mathrm{out}}) & \nonumber\\
  \exp \biggl[ \frac{i}{\hbar} \biggl(
    \int\limits_{\mathbf{r}_{\mathrm{trap},n}}^{\mathbf{r}} \mathbf{p}
    \,\mathrm{d}\mathbf{r}' \biggr) - \int_{0}^{t} \hbar\omega_n
    \,\mathrm{d}t' \biggr]
  \end{eqnarray}
where $\omega_n$ is the temporal de Broglie frequency of each atom laser
defined as $\hbar\omega_n = mgz_{\mathrm{trap},n} + \mu_n(t')$. This
frequency is the sum of the initial atomic kinetic energy and $n$-th
condensate's chemical potential $\mu_{n}$, where $m$ denotes the atom mass
and $g$ the gravitational acceleration. We stress that the temporal phase
fluctuations of both the initial condensates and the outcoupling process are
accounted for by assuming time-dependent chemical potentials $\mu_n(t')$.
After outcoupling, due to energy conservation $\hbar\omega_n$ remains
constant, i.e. $\hbar\omega_n = mgz + \mu_n(t_{\mathrm{out}})$ for $t' >
t_{\mathrm{out}}$ (with $0 \le t_{\mathrm{out}} \le t$). We are interested
in the total wavefunction outcoupled from the $N$ sources all emitting at
the same vertical position $z_{\mathrm{trap},n} = z_{\mathrm{trap}}$,
$y_{\mathrm{trap},n} = 0$, and equally spaced along the $x$ axis with
$x_{\mathrm{trap},n} = d \cdot n$, where $d = \lambda_{\mathrm{CO_2}}/2$ is
the lattice spacing. In the far field one finds $p_{z,n}= \sqrt{2m(mg
(z_{\mathrm{trap}}-z)- \mu_n(t_{\mathrm{out}}))-(p_x^2+p_y^2)}$ and $p_{x}=
mx/t_{\mathrm{exp}}$, where $t_{\mathrm{exp}}=t-t_{\mathrm{out}}$ denotes
the free expansion time. After carrying out the above integrals, we find
that the probability to detect a particle at fixed time $t$ and position
$\mathbf{r}$ can be written as:
\begin{equation}
  \bigl|\psi(\mathbf{r}, t)\bigr|^2 =
  \bigl|C(\mathbf{p}, t_{\mathrm{out}})\bigr|^2
  \sum_{n=1}^{N} A_n \cos \left[ \frac{ndm}{\hbar t_{\mathrm{exp}}}\,x + 
    \delta_n + \beta_n(z,t) \right]
  ~,
\end{equation}
where $A_0 = \sum_{m=1}^{N} a_m^2$, $\delta_0 = 0$ and for $1 \le
n \le N-1$: $A_n e^{i\delta_n} = 2 \sum_{m=1}^{N-n} a_m a_{m+n}
  e^{i\Delta\theta_m}$. Both $A_n$ and the phase angles $\delta_n$ shall
here be real numbers. These coefficients can be evaluated in a random
walk model with $N-1$ steps assuming that the phase angles
$\Delta\theta_m=\theta_m-\theta_{m+n}$ are randomly
distributed~\cite{20}. If all amplitudes $a_m$ are equal,
for $N\gg1$ the average fringe visibility $V\equiv A_1/A_0$ is given by
$\sqrt{\pi/N}$.
The $z$-dependent phase in eq.~2 is
\begin{equation}
  \beta_n(z,t) = \frac{1}{\sum_{m'=1}^{N-n} e^{i\Delta\theta_m'}}
  \cdot\sum_{m=1}^{N-n} \bigl[ \varphi_m(z,t) - \varphi_{m+n}(z,t) \bigr]
  e^{i\Delta\theta_m}
  ~,
\end{equation}
which again can be evaluated with a random walk model. In this
formula
$\varphi_n(z,t)=(1/\hbar)\int^{t_{\mathrm{out}}(z)}_{0}{\mu_{n}(t')dt'}+\mu_{n}(t_{\mathrm{out}}(z))(t-t_{\mathrm{out}}(z))$.

We expect that the mean square value of this $z$-dependent phase
is approximately given by $\sqrt{\langle\Delta\beta_n^2\rangle}
\approx \sqrt{\langle\Delta\varphi^2\rangle}$, which is
independent of the number of condensates.  Small statistical
fluctuations of the chemical potential, which would lead to a line
broadening, should affect the multiple beams interference pattern in a
similar amount than a two atom lasers interference experiment. In
case that the lattice axis is tilted against the axis of gravity
by an angle $\alpha$ (with $\alpha \ll1$), the cosine factor in
eq.~2 for $|\psi|^2$ contains an additional factor
$\frac{nd}{\hbar} \sin(\alpha) \sqrt{2m^2g(z_{\mathrm{trap}}-z)}$,
which in turn will lead to a tilting of the fringe pattern.

The experimental interference pattern shown in
Fig.~\ref{fig:pattern} has high contrast (near 50\,\%). With our
present imaging resolution of $6\,\mu\mathrm{m}$, we spatially
resolve only the interference pattern between adjacent lattice
sites. Let us begin an analysis of the atom lasers interference
pattern by restricting ourselves to signal profiles for a given
fall distance $|z - z_{\mathrm{trap}}|$, which corresponds to a
fixed outcoupling time $t_{\mathrm{out}}$. In this case, the amplitude and
phase of the interference pattern are evaluated analogous to
earlier work on the interference of independent Bose-Einstein
condensates~\cite{22,23}. The average fringe contrast of a set of
eight observed interference patterns with its corresponding
standard deviation was $(24 \pm 5)\,\%$. Within the quoted
uncertainties, this value agrees very well with the results of our
previous works studying the interference of a variable number $N$
of Bose-Einstein condensates~\cite{17}, from which we interpolate
a value of 25\,\% for $N = 7$. From this we conclude that the atom
laser output coupling mechanism, within our experimental
measurement uncertainties, does not introduce high frequency phase
noise and is reproducible from site to site.

The present work involves atom lasers with (quasi-)continuous output coupling,
and we can study the variation of the phase of the interference pattern with
vertical position $z$, which directly is related to the duration of free fall.
This allows us to monitor the atom lasers relative frequency. The vertical,
stripe-like nature of the fringes shown in Fig.~\ref{fig:pattern} clearly
shows that the temporal evolution of the atom laser phase over the observed
2\,ms time window is well controlled.  Fluctuations of the relative de Broglie
frequencies here would lead to jittering or blurred fringes.

From Eq. 2 we expect that the fringe spacing increases linearly
with the expansion time $t_{\mathrm{exp}}$ of atoms and is given
by $h t_{\mathrm{exp}}/(md)$, where $d$ is the spatial separation
of sources. As the atoms during the outcoupling process already
transversally expand before being completely removed from the trap
potential, we set $t_{\mathrm{exp}} = t_{\mathrm{fall}} +
t_{\mathrm{delay}}$ with $z = z_{\mathrm{trap}} -
gt_{\mathrm{fall}}^2/2$ describing the ballistic free fall and a
constant term $t_{\mathrm{delay}}$ to account for the time lag
during outcoupling in which atoms are not yet exposed to the
earth's gravitational field alone. A fit to the experimental data
for the pattern of Fig.~\ref{fig:pattern} gives
$t_{\mathrm{delay}} \approx 3\,\mathrm{ms}$, which is consistent
with a simple numerical simulation. Fig.~\ref{fig:fringes}a shows
the experimental fringe spacing along with the fit (solid line) as
a function of the expansion time.

Of particular interest is an analysis of the variation of the phase of the
fringe pattern on the expansion time. The corresponding data is shown in
Fig.~\ref{fig:fringes}b. Over the observed 2\,ms time interval, corresponding
here to roughly $250\,\mu\mathrm{m}$ of fall distance, the phase remains
relatively constant with in average showing variations of order of~$2\pi/10$.
The experimental data can both be fitted with a constant (dashed line) and
linear (solid line) function of fall time, with the latter fit yielding a
smaller sum of residuals. The non-zero slope of the latter fit translates to a
finite difference frequency between adjacent atom lasers, as is e.\,g.\ 
expected for a CO$_2$-laser lattice beam non-perfectly aligned orthogonal to
the axis of gravity. Table~\ref{table} gives the fitted average atom laser
difference frequency for this measurement (data set~\#1) along with the result
for three other high contrast fringe patterns. The average difference
frequency between adjacent atom laser beams for the four data sets equals
$(-51 \pm 46)\,\mathrm{Hz}$.  Notably, this result corresponds to the
difference frequency of independently generated matter wave beams.

To conclude, we have observed the interference of an array of atom laser
sources.  The average relative phase fluctuations and difference frequencies
between adjacent matter wave sources were monitored. For the future, we expect
that the relative measurement of atom laser frequencies can allow for novel
matter wave metrology techniques.  An intriguing question is if ultimately
frequency measurements of small atomic mass differences can be carried out.

\begin{acknowledgments}
  We acknowledge financial support from the Deutsche Forschungsgemeinschaft,
  the Landesstiftung Baden W\"urttemberg, and the European Community.
\end{acknowledgments}

\small{}

\clearpage
\begin{figure}
\caption{Scheme of the periodic trapping potential during the atom
lasers emission. The emission occurs at the downwards directed
surfaces of the microscopic traps, as indicated in the figure.}
\label{fig:scheme}
\end{figure}

\begin{figure}
\caption{Far field interference pattern of independent all-optical atom
  lasers. (a) Absorption image of the interference pattern of seven overlapped atom lasers beams,
  recorded at an expansion time of 15\,ms after the onset of outcoupling of
  atoms from the sites of the optical lattice. The field of view is $60\,\mu\mathrm{m} \times
  220\,\mu\mathrm{m}$. (b) Horizontal density profile of
  image (a) averaged over the marked vertical region with $44\,\mu\mathrm{m}$
  height. The experimental data is represented by connected solid dots and the
  fitted fringe pattern by the dashed line.}
\label{fig:pattern}
\end{figure}

\begin{figure}
\caption{Fringe spacing and phase for the interference pattern of
  Fig.~\ref{fig:pattern}a as function of the expansion time and of the
  corresponding free fall distance. (a) Experimental data for the fringe
  spacing evolution along with a linear fit (solid line). (b) Evolution
  of the fringe phase. The data has been fitted both
  with a constant (dashed line) and a linear function (solid line).
  The linear fit gives a finite slope which
  corresponds to an (average) difference frequency between adjacent atom
  lasers of $-77\,\mathrm{Hz}$.}
\label{fig:fringes}
\end{figure}

\begin{table}
  \begin{tabular}{|c|c|}
    \hline Data set number & $\delta\omega_{n,n-1}$ (Hz) \\\hline
    1 & $-77$ \\\hline
    2 & $-75$ \\\hline
    3 & $-133$ \\\hline
    4 & $82$ \\\hline
    Average & $-51 \pm 46$ \\\hline
  \end{tabular}
  \caption{Measured (averaged) difference frequency of neighbouring atom laser
    beams. From a total of eight recorded interference patterns, with, due to
    the intrinsically random nature of the individual condensates phases,
    statistically distributed fringe contrast, these four sets had high enough
    contrast to allow for a reliable extraction of the average relative de Broglie
    frequency. We have fitted the measured differential phase with a linear
    function of the expansion time, as shown in Fig.~\ref{fig:fringes}b for
    data set \#1. In the bottom row, the mean value for the matter wave
    frequency along with its standard deviation is given.}
\label{table}
\end{table}

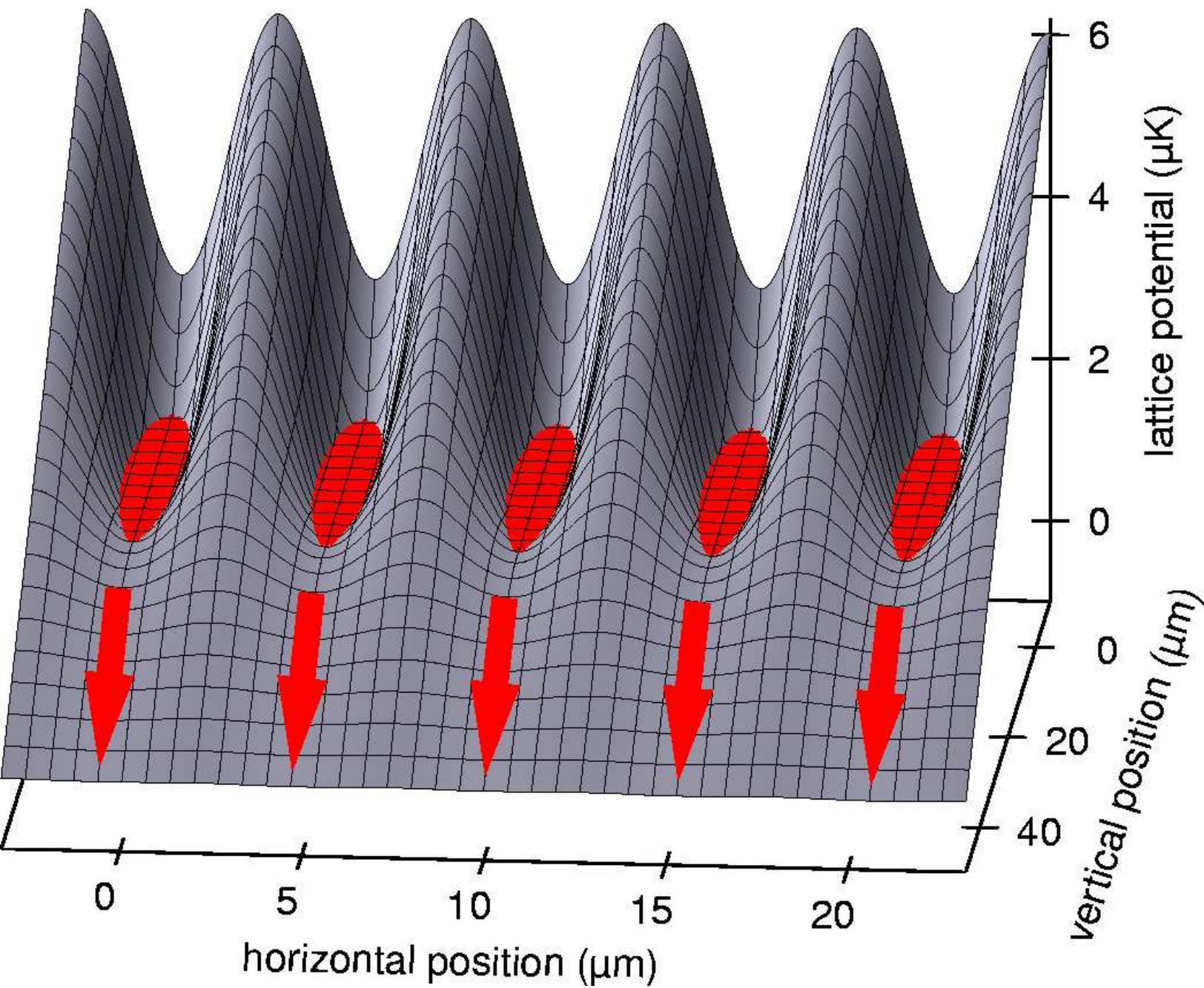
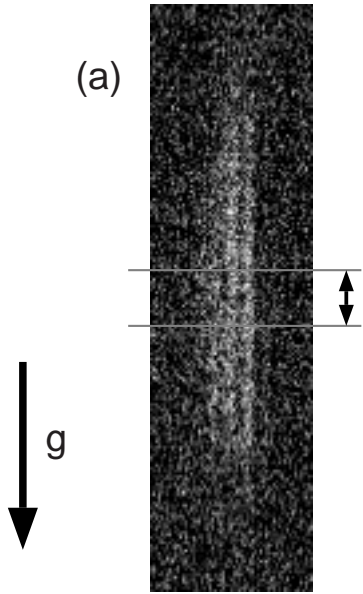
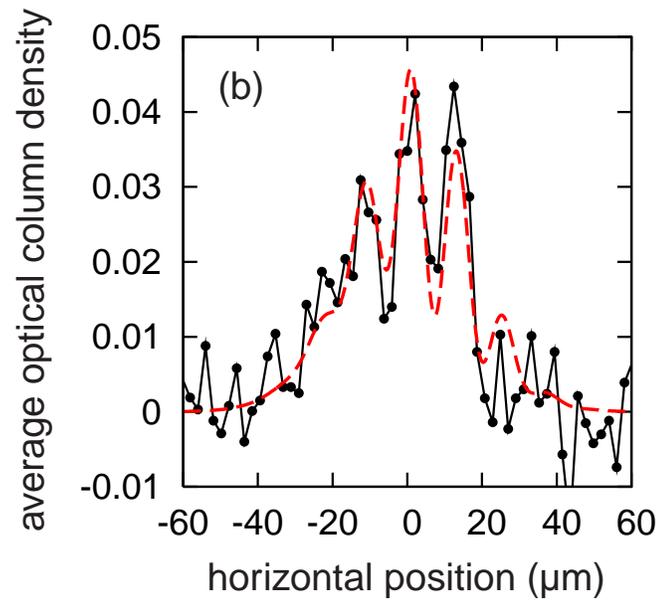
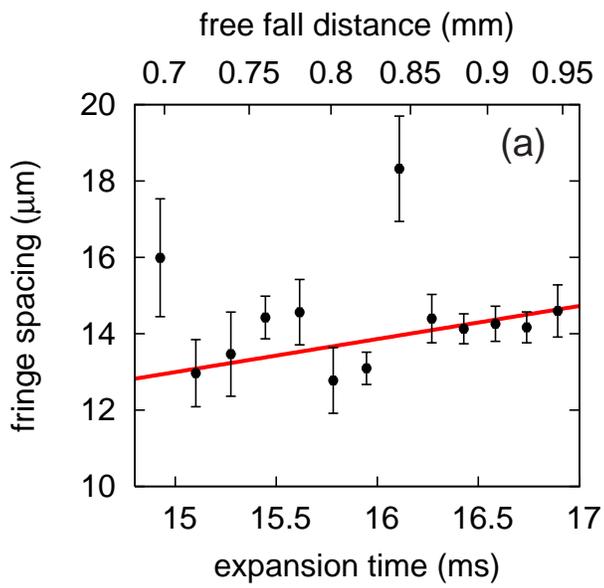
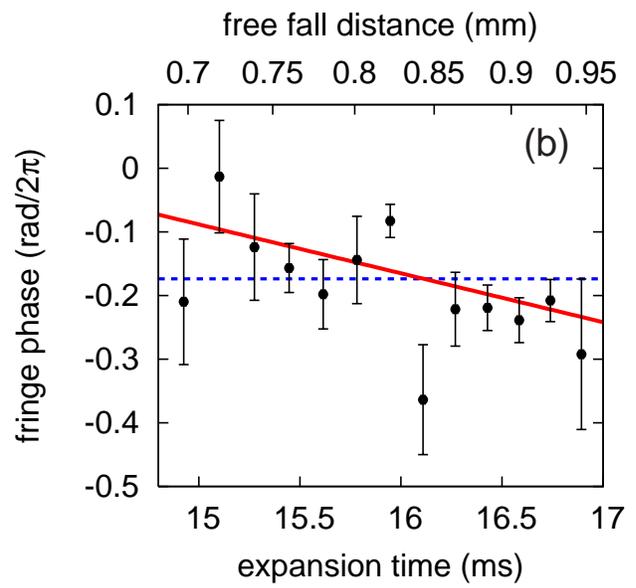
\end{document}